\documentclass[doublecol,amsmath,amssymb]{epl2}

\usepackage{epsf}
\usepackage{rotating}
\usepackage{amsmath,amsfonts}
\usepackage[dvips]{psfrag}

\usepackage{graphicx}
\usepackage{dcolumn}
\usepackage{bm}

\def\pmb#1{\setbox0=\hbox{#1}%
  \kern-.025em\copy0\kern-\wd0
  \kern.05em\copy0\kern-\wd0
  \kern-.025em\raise.0433em\box0}

\def\bfnabla{\mbox{\boldmath $\nabla$}}
\def\bfe{\mbox{\bf e}}

\def\bfr{\mbox{\bf r}}

\def\bfu{\mbox{\bf u}}
\def\bfB{\mbox{\bf B}}

\newcount\eqnno
\newcount\secno   
\newcount\subno   

\def\section#1
   {\vskip0pt plus.1\vsize\penalty-250
    \vskip0pt plus-.1\vsize\vskip18pt plus9pt minus6pt
    \subno=0 \eqnno=0 
    \global\advance\secno by 1
   \centerline{\bf\the\secno. #1}
    \medskip}

\def\subsection#1
   {\vskip-\lastskip
        \vskip15pt plus4pt minus4pt
    \bigbreak
    \global\advance\subno by 1
    \centerline{\it \the\secno.\the\subno. #1}\smallskip}

\def\sen{\the\secno}

\def\eqn#1{\global\advance\eqnno by 1
           \eqno(\sen.\the\eqnno)
           \expandafter \xdef\csname #1\endcsname
           {\sen.\the\eqnno}\relax }
\def\eqnp#1#2{\global\advance\eqnno by 1
           \eqno(\sen.\the\eqnno\hbox{#2})
          \expandafter \xdef\csname #1\endcsname
           {\sen.\the\eqnno}\relax }
\def\eqnpr#1{\eqno(\sen.\the\eqnno\hbox{#1})}
\def\eqnn#1{\global\advance\eqnno by 1
           (\sen.\the\eqnno)
           \expandafter \xdef\csname #1\endcsname
           {\sen.\the\eqnno}\relax }
\def\eqnm#1#2{\global\advance\eqnno by 1
           (\sen.\the\eqnno\hbox{#2})
           \expandafter \xdef\csname #1\endcsname
           {\sen.\the\eqnno}\relax }
\def\eqnr#1{(\sen.\the\eqnno\hbox{#1})}

\catcode`\£=\active \gdef£{\setbox0=\hbox{0}\hbox to\wd0{}}
\newsavebox{\thalfbox}
\sbox{\thalfbox}{$\textstyle\frac{1}{2}$}

\newsavebox{\shalfbox}
\sbox{\shalfbox}{$\scriptstyle\frac{1}{2}$}

\newsavebox{\squartbox}
\sbox{\squartbox}{$\frac{1}{4}$} 

\newsavebox{\etbox}
\sbox{\etbox}{\boldmath$\eta$}

\newsavebox{\astrutbox}
\sbox{\astrutbox}{\rule[-5pt]{0pt}{20pt}}

\title{Relations Between Dynamo-Region Geometry 
and the Magnetic Behavior of Stars and Planets}

\author{Laure Goudard \and Emmanuel Dormy}
\shortauthor{Laure Goudard \& Emmanuel Dormy}
\institute{MAG (IPGP \& ENS), CNRS UMR7154, LRA D\'epartement de Physique, 
24 rue Lhomond, 75005 Paris France.}


\abstract{The geo and solar magnetic fields have long been thought to be
  very different objects both in terms of spatial structure and temporal
  behavior. The recently discovered field structure of a fully convective
  star is more reminiscent of planetary magnetic fields than the Sun's
  magnetic field \cite{BIB1}, despite the fact that the physical and chemical
  properties of these objects clearly differ. This observation suggests
  that a simple controlling parameter could be responsible for these
  different behaviors. We report here the results of three-dimensional
  simulations which show that varying the aspect ratio of the active dynamo
  region can yield sharp transition from Earth-like steady
  dynamos to Sun-like dynamo waves.}

\begin{document}

\maketitle

\noindent {\bf Introduction.}\! --
Observations of the magnetic fields due to dynamo activity appear to fall
into two categories: fields dominated by large-scale dipoles (such as the
Earth and a fully convective star), and fields 
{whith smaller-scale and 
non-axisymmetric structures} (such as the Sun). Moreover two
kinds of 
different temporal behaviour have been identified so far: very irregular
polarity reversals (as in the Earth), and quasi-periodic reversals (as in
the Sun). Since the Earth and the Sun provide the largest database of
magnetic field observations, these objects have been well studied and
described in terms of alternative physical mechanisms: the geodynamo
involves a steady branch of the dynamo equations, perturbed by strong
fluctuations that can trigger polarity reversals, whereas the solar dynamo
takes the form of a propagating dynamo wave. The signature of this wave at
the Sun's surface yields the well-known butterfly-diagram (Sunspots
preferentially emerge at a latitude that is decreasing with time during the
solar cycle).\\

\noindent {\bf Modelling.}\! --
Because of their very different natures (liquid metal in one case, plasma
in the other), planetary and stellar magnetic fields are studied by
different communities.
Non-dimensional numbers controling the dynamics of the Earth and the Sun,
for example, do significantly differ (see~\cite{Zhang,lelivre}).
As a practical matter however, the techniques 
{as well as the typical parameters}
used in numerical studies of these two systems are surprisingly similar. 
To some
extent this is due to the restricted parameter space available to present
day computations. 
{The parameter regime numerically accessible 
is rather remote from the actual objects. For planetary dynamos the 
main discrepancy relies in the rapid rotation in the momentum equation 
(characterized by the {\it Ekman number}), whilst 
for stellar dynamos it relies in solving the induction equation 
with weak resistive effects (characterized by high values of the 
{\it magnetic Reynolds number}).}
Yet within this restricted domain, the sharply different
key characters to both geo \cite{BIB2} and solar \cite{BIB3,BIB4} magnetic 
fields have been reproduced. This leads us to argue that the important 
parameter controlling
the magnetic field behaviour is the aspect ratio of the dynamo region
(i.e. the radius ratio of the inner bounding sphere to the outer bounding
sphere). Indeed, in the Earth, the inert solid inner core extends to less
than 40\% of the core radius, whereas in the Sun, the radiative zone fills
70\% of the solar radius. One expects the convective zones of stars and
planets to have all possible intermediate aspect ratios, even extending
to fully convective spheres. 

\begin{figure*}
\centerline{\includegraphics[width=15cm]{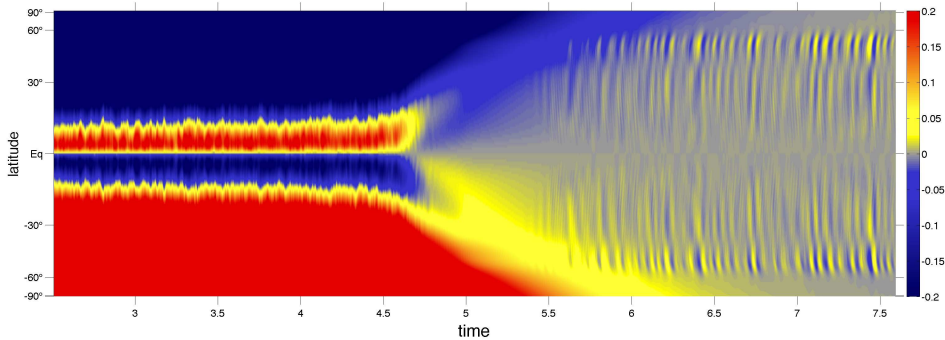}}
\caption{Time evolution of the radial magnetic field averaged in longitude
(for an aspect ratio of $0.65$). The initial dipole field survives for a few
diffusion times, and then vanishes to yield a butterfly-like
diagram.}
\end{figure*}

In order to isolate and understand this
purely geometrical effect, we have carried out three-dimensional
numerical simulations of self-excited convective dynamos in which the
domain aspect ratio was slowly varied, with all other parameters held
constant. The governing equations as well as parameter regimes used
here were originally introduced for a geodynamo reference calculation
\cite{BIB5}. The only distinction being the use of stress-free boundary
conditions on the outer sphere of the domain, while imposing no-slip
boundary conditions at the bottom of the convective region. This choice
was made in order to create a strong shear at the base of the model, 
and thus try to mimic the solar tachocline \cite{BIB6}. The inner sphere is
here assumed to be insulating, and we use differential heating.
The governing equations are in non-dimensional form:
\begin{eqnarray}
{\rm E} \, 
\left[\partial _t \bfu +  (\bfu \cdot \bfnabla) \bfu \right]
\!\!\!\!&=& \!\!\!\!
- \bfnabla \pi 
+ {\rm E} \, \Delta u
- 2 \bfe_z \times \bfu \nonumber\\
\!\!\!\!&+&\!\!\!\! \widetilde{Ra} \, \bfr \, \theta
+{\rm Pm}^{-1} \left(\bfnabla \times \bfB \right) \times \bfB\, ,
\end{eqnarray}   
\begin{equation}
\partial _t \bfB = \bfnabla \times (\bfu \times \bfB) 
+ {\rm Pm}^{-1} \, \Delta \bfB \, ,
\label{induction}
\end{equation} 
\begin{equation}
\partial_t \theta + (\bfu \cdot \bfnabla) (\theta+T_s) 
= {\rm Pr}^{-1} \Delta \theta\, ,
\end{equation} 
\begin{equation}
\bfnabla \cdot \bfu =  \bfnabla \cdot \bfB = 0 \, ,
\end{equation} 
where
\begin{equation}
{\rm E}=\frac{\nu}{\Omega D^2}\, , \ 
\widetilde{Ra}=\frac{\alpha g \Delta T D}{\nu \Omega}\, , \ 
{\rm Pr}=\frac{\nu}{\kappa}\, , \ 
{\rm Pm}=\frac{\nu}{\eta}\, .
\end{equation} 
All simulations reported here were performed keeping the following
parameters constant
${\rm E}=10^{-3}$, $\widetilde{Ra}=100$, ${\rm Pr}=1$, ${\rm Pm}=5$~.
The above system is integrated in three--dimensions of space (3D)
using the Parody code \cite{Parody}.

When the inner (non dynamo generating) body occupies
less than about 60\% of the convective body in radius, the flow generates a
dipolar field, very similar to that of the Earth. It features patches of
intense flux at high latitudes and some reversed patches at low latitude,
similar to the ones revealed by a downward continuation of the Earth's
field to the Core-Mantle boundary \cite{BIB7}. This strongly dipolar solution
becomes unstable with a further increase of the aspect ratio. For an aspect
ratio of $0.65$ --close to that of the Sun-- the strong dipole is first
maintained and then {strongly weakens}, but dynamo action continues in a
different form: that of a wavy solution with quasi-periodic reversals
(Fig. 1), reminiscent of {some aspects of} the solar magnetic
field {behavior}. Drifting features can be
observed both on the radial field at the surface of the model (Fig. 1 \& 2b)
and on the azimuthal (east-west) field below the surface of the model
(Fig. 2c). Due to the complex nature of these fully tri-dimensional
simulations, many waves can co-exist. Some of the dominant structures
appear to propagate toward the equator; others propagate poleward. 
Reversed waves are also observed at the surface of the Sun at higher
latitudes \cite{BIB8}.
{Let us stress however that the model cannot be expected to
capture all the features either of the geo or solar magnetic fields.
In particular due to the parameters regime and the lack of stratification
in our modelling.}
\\ 

\noindent {\bf Physical interpretation.}\! --
In order to investigate the physical mechanisms associated
with these waves, we have performed some kinematic simulations. During the
course of the simulation the Lorentz force was suppressed. The wavy nature
of the dynamo field was unaltered by this modification. This rules out the
possibility of an interpretation in terms of pure Alfv\`en waves or
Alfv\`en waves modified by rotation (so called MC or MAC waves), which
both require the back-reaction of the Lorentz force. Of course, suppressing
the Lorentz force is not without consequences: the flow slowly evolves to a
different purely hydrodynamical state, and the magnetic field now grows
exponentially, but both of these effects are sufficiently slow for the
wavelike character to persist over many wave periods. 

Two other
interpretations for the nature of these waves remain possible: either
hydrodynamic fluctuations (e.g. inertial waves or Rossby waves) or dynamo
waves, as expected on the Sun. These possibilities were tested by comparing
oscillations in the velocity field and in the magnetic field in the
kinematic simulations. We found that a high frequency signal is present
both in the flow and in the magnetic field. This demonstrates the presence
of hydrodynamic waves, which induce magnetic fluctuations. The lower
frequency signal is however absent in the flow. This provides a proof of
their ``dynamo wave'' nature.  

\begin{figure}
\centerline{\includegraphics[width=9.2cm]{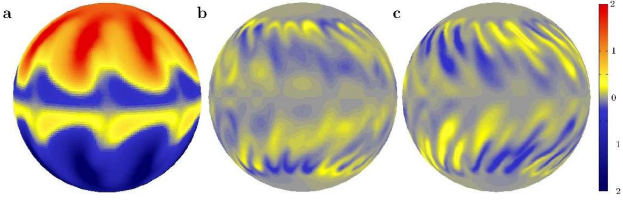}}
\caption{Radial magnetic field at the surface of the outer sphere,
for aspect ratios of $0.45$ (a) and $0.65$ (b). Azimuthal magnetic field below
the surface of the $0.65$ aspect ratio model (c).}
\end{figure}

\begin{figure}
\centerline{\includegraphics[width=7.6cm]{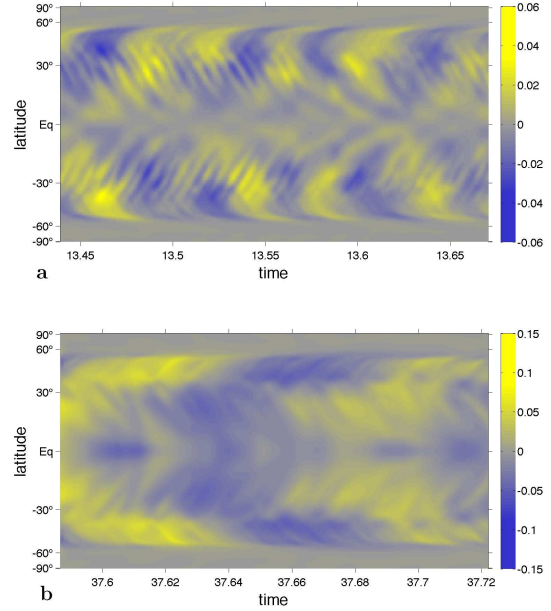}}
\caption{Time evolution of
the {zonal average of the} azimuthal magnetic field below the surface of the
model, for an aspect ratio of $0.65$: the antisymmetric (a) and symmetric (b)
solutions.}
\end{figure}

We have numerically observed such dynamo waves for aspects ratio up to
$0.8$. {For the parameters investigated here, the}
transition from a dynamo dominated by a fluctuating dipole to a
dynamo wave occurs for an aspect ratio close to $0.65$. This transition
exhibits hysteresis: once a dynamo-wave solution is present, the aspect
ratio can be reduced again down to $0.6$, while maintaining this dynamo
mode.\\

\noindent {\bf Connections with parameterized models.}\! --
Butterfly diagrams indicative of the solar cycle are usually produced using
simplified parameterized models or ``mean field'' models. These models
require a prescription of the turbulent induction, the so-called 
``$\alpha$--effect'' (which can also be introduced in terms of
deviation from axisymmetry \cite{Braginsky}).
{We should stress that this is a valid 
approximation only if certain conditions are satisfied (e.g. \cite{HUGHES}).}
Such butterfly--like
diagrams are generally not produced by direct three-dimensional modelling,
with the notable exception (only in the reverse direction) of the
pioneering work of Gilman and Glatzmaier \cite{BIB3,BIB9}. 

Because of the strong symmetry of the convective flows influenced by the
rapid rotation of the planet or the star, it is well known that two
independent families of solutions exist, namely with dipole symmetry
(antisymmetric with respect to the equator) and
quadrupole symmetry (symmetric with respect to the equator). 
Both families of solutions are often described in reduced
parameterised models \cite{BIB10,BIB11}, and we have observed these two
families in our fully 3D simulations (Fig.~3). Both branches are stable 
in our simulations for
long periods of time, but can also be destabilised to yield a change of
symmetry. In fact, despite the relative complexity of our model, the
temporal behavior of both symmetries is clearly reminiscent of kinematic
studies of earlier reduced models (Fig.~4c,d and \cite{BIB10}).  

The simpler meanfield equations for the axisymmetric field 
are obtained by writing the flow and field as
\begin{equation}
\bfu= s\, \omega\, \bfe_\phi\, , \qquad
\bfB=\bfB_p + B \,\bfe_\phi
=\bfnabla \times (A \, \bfe_\phi) +B \,\bfe_\phi \, ,
\end{equation} 
i.e. assuming a mean flow in the form 
of a zonal shear only.
In the isotropic case,
the axisymmetric part of (\ref{induction}) yields
(e.g. \cite{lelivre})
\begin{equation}
\frac{\partial A}{\partial t}=\alpha B + {\rm Rm}^{-1} \, {\cal D}_2 A \, ,
\label{alpha1}
\end{equation} 
\begin{equation}
\frac{\partial B}{\partial t}= s \, \bfB_p \cdot \bfnabla \omega
+ \left(\bfnabla \times \alpha \bfB_p  \right) \cdot \bfe_\phi
+ {\rm Rm}^{-1} \,{\cal D}_2 B  \, ,
\label{alpha2}
\end{equation} 
where $s$ denotes the cylindrical radius and
${\cal D}_2 = \Delta - {1}/{s^2}$ (note that Pm in (\ref{induction}) is here changed
to Rm as the flow is now assumed to be given).

For an instability of (\ref{alpha1}-\ref{alpha2}) to exist, these equations
must not decouple (this is the essence of Cowling's anti-dynamo theorem \cite{Cowling}).
Equation (\ref{alpha2}) involves $A$ through two terms. Reduced models have
been classified in two categories depending on the dominant term. 
The first term on the RHS of (\ref{alpha2}) involves the zonal shear and is
referred to as the $\Omega$--effect. The second term in the RHS of
(\ref{alpha2}), as well as the first term on the RHS of (\ref{alpha1}), involve
mean induction from non-axisymmetric features in the flow and are referred
to as the $\alpha$--effect. 

Dropping the $\alpha$--effect term in (\ref{alpha2}) and
writing the resulting equations in a simplified cartesian
geometry yields
\begin{equation}
\frac{\partial A}{\partial t}=\alpha B + {\rm Rm}^{-1} \, \Delta A \, ,
\quad
\frac{\partial B}{\partial t}= G \, \frac{\partial A}{\partial x}
+ {\rm Rm}^{-1} \,\Delta B  \, ,
\label{alpha34}
\end{equation} 
where $G = {\rm d}u_y / {\rm d}z$.
Parker \cite{Parker55} was the first to identify travelling waves solutions
(dynamo waves) of the above system. 
These oscillatory dynamos, named Parker waves, were obtained
by Roberts \cite{BIB10} for nearly axisymmetric dynamos in spherical
geometries (following the formalism of Braginsky \cite{Braginsky}). 
It was found that while the $\alpha\Omega$--dynamos tended  to be 
oscillatory (complex growth rate), for $\alpha^2$--dynamos 
the simplest dipole solutions tended to be stationary (real growth rate).
A similar behavior can easily be traced in the simpler cartesian example
above (see also \cite{Moffatt,Parker79} for a discussion of the
generic behavior of such nearly axisymmetric mean field dynamos).

\begin{figure*}
\centerline{\includegraphics[width=14cm]{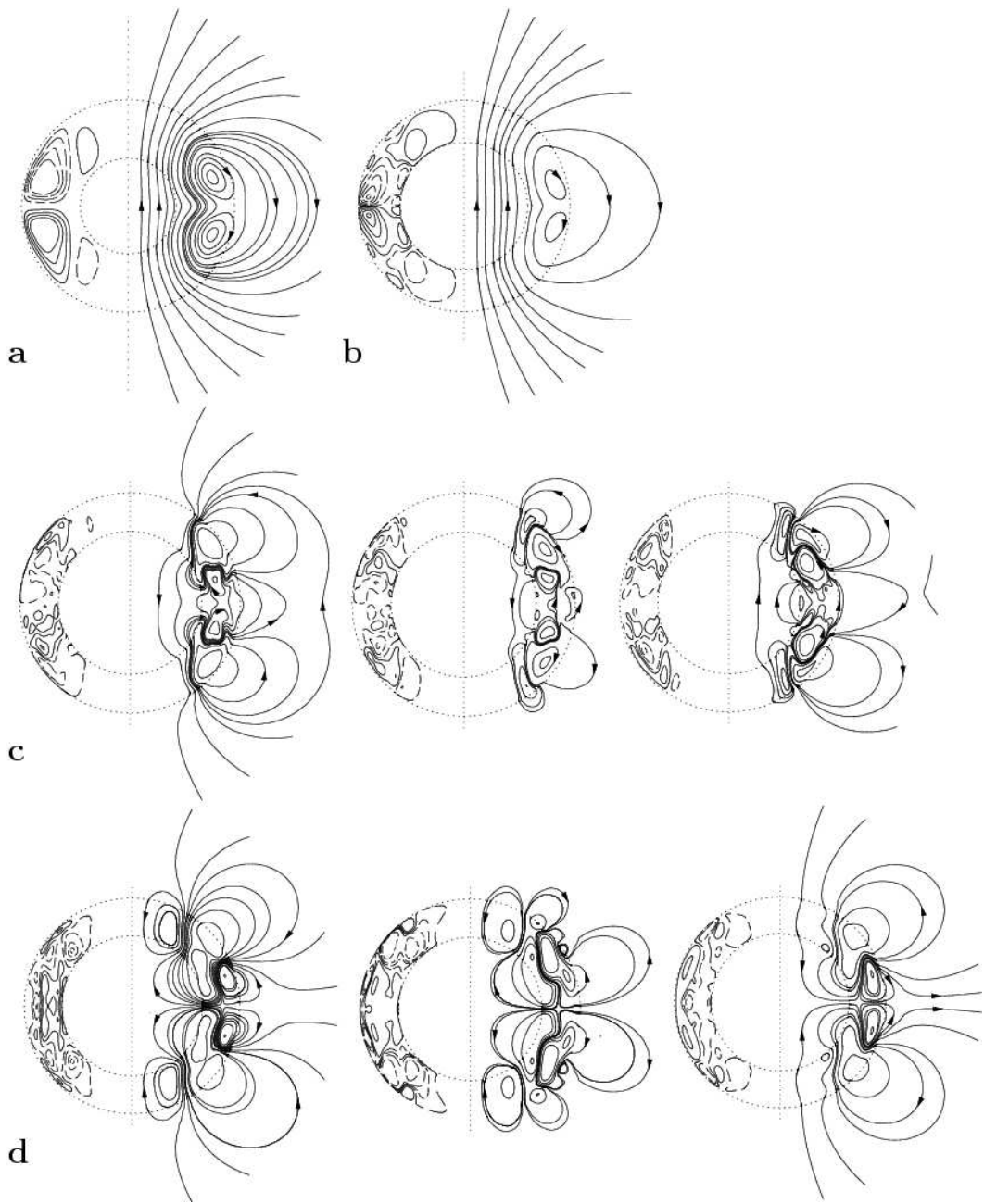}}
\caption{The {zonal average of the } magnetic field in our 3D simulations.
Contours of the toroidal (east-west) part of the field are
plotted in the left hemisphere and lines of force of the meridional
(poloidal) part of the field plotted in the right hemisphere. 
The aspect
ratio is increased from $0.45$ (a) to $0.6$ (b) and to $0.65$ (c-d). The sequence
of dynamo waves is represented for the antisymmetric mode (c) and symmetric
mode (d). It is similar in nature to that produced by parameterized models
\cite{BIB10}.}
\end{figure*}

We can perform further comparisons with reduced models by studying only
the axisymmetric component of the simulated field. Figure~4 shows the 
{azimuthally}
averaged field for some of our fully 3D simulations.  The Earth-like mode is
represented for aspect ratios of $0.45$ and $0.6$ (a \& b). The active dynamo
region lies outside the tangent cylinder \cite{BIB2}, it therefore gets
increasingly constrained as the inner-sphere in increased. The dipole
eventually drops for large aspect ratio, when the volume outside the
tangent cylinder becomes too small.
Weakly dipolar solutions were also obtained at large aspect ratio
in simulations using equations modified by hyperviscosity
\cite{Heimpel}. The dipolar solution was also found to decay and
eventually vanish by increasing the aspect ratio in a 
reduced parameterized model for the Earth's core 
\cite{BIB12}. 
Here we show that the steady dynamo branch can be replaced,
at larger aspect ratio, by an oscillatory dynamo mode.
Comparison with reduced parameterized models can help
interpret this transition to the solar-like mode. 
A strong zonal wind develops, in our simulations, in the Solar-like
mode. Although the terminology 
of parameterized models must be used with care for direct simulations (the
hypothesis of scales separation does not strictly apply), this suggests a
transition from a dynamo of the $\alpha^2$ type to a dynamo of the $\alpha
\Omega$ type as the 
aspect ratio is increased. Indeed, Earth-like three-dimensional models have
been interpreted in terms of regeneration by convective vortices only, and
thus closer to the $\alpha^2$ formalism \cite{BIB13} (sometimes referred to
as ``giant $\alpha$--effect''), whereas the $\alpha \Omega$ formalism
provides the classical framework to model solar dynamo-waves, as guided by
the strong shear at the base of the convection zone
\cite{BIB10,BIB11}. Such nearly axisymmetric
dynamos \cite{BIB10} produce cyclic magnetic behaviour very similar to
the cycles examplified on Figure~3.\\ 

\noindent {\bf Conclusions.}\! --
By varying the aspect ratio, we have observed a sharp transition from
a dipole dominated large scale-magnetic field 
{to a cyclic dynamo with a weaker dipole.}
This indicates that the geometry of the dynamo region severly constrains 
the existence of the dipole dominated solution. We should however stress
that other parameters, involving
ratio of typical forces, could affect the precise value of the critical 
aspect ratio for transition. 
{The values of these parameters in our simulations (as in 
all numerical models to date) are indeed
very remote from the actual relevant values for the Sun, or for the Earth.
The potentially strong effect of this parameter change on the dynamo
solution should not be under-estimated. It is indeed quite striking, that
despite these shortcomings, numerical models can capture a good part of the
qualitative feathures of the solar and geo-magnetic fields.}

Recent observations of stellar magnetism appear to 
corroborate this mechanism. Donati {\it et al.} \cite{BIB1} reported 
observations of a strongly dipolar field in a fully convective star (V374
Peg).
More recently, Donati {\it et al.} \cite{Donati2} report magnetic observations
of $\tau$--Bootis, a rapidly rotating F star, i.e. one with a relatively 
shallow outer convection zone.  Not only did they observe a rather complex 
magnetic field structure, but they also report that the 
overall polarity of the magnetic field has reversed after one year of
observation. They interpreted this observation as an indication that
the large aspect ratio $\tau$--Bootis star is undergoing magnetic cycles,
similar to those of the Sun. 

Futher observations of planets and stars are needed, but clearly the
observations available so far seem to confirm the important role of the
aspect ratio in controling the transition from steady to cyclic dynamo
modes.

\acknowledgments
The computing resources for this work were provided by the CNRS-IDRIS, the
ENS-CEMAG and the IPGP-SCP computing centers. 
We are grateful to Pr. N.O.Weiss for discussions on a preliminary version
of this work.

\end{document}